# Spin wave eigenmodes in nanoscale magnetic tunnel junctions with perpendicular magnetic anisotropy

Andrea Meo[1], Chengcen Sha[2], Emily Darwin[3], Riccardo Tomasello[1], Mario Carpentieri[1], Ilya N. Krivorotov[2], Giovanni Finocchio[4]

[1] *Department of Electrical and Information Engineering, Politecnico di Bari, I-70125 Bari, Italy*

[2] *Department of Physics and Astronomy, University of California, Irvine, California 92697, USA*

[3] *Empa, Swiss Federal Laboratories for Materials Science and Technology, CH-8600 Dübendorf, Switzerland*

[4] *Department of Mathematical and Computer Sciences, Physical Sciences and Earth Sciences, University of Messina, I-98166 Messina, Italy*

[*]Corresponding author: giovanni.finocchio@unime.it

**Abstract**

Magnetic tunnel junctions (MTJs) are key enablers of spintronic technologies used in a variety of applications including information storage, microwave generation and detection, as well as unconventional computing. Here, we present experimental and theoretical studies of quantized spin wave eigenmodes in perpendicular MTJs focusing on a coupled magnetization dynamics in the free (FL) and reference (RL) layers of the MTJ, where the RL is a synthetic antiferromagnet (SAF). Spin-torque ferromagnetic resonance (ST-FMR) measurements reveal excitation of two spin wave eigenmodes in response to applied microwave current. These modes show opposite frequency shifts as a function of out-of-plane magnetic field. Our micromagnetic simulations accurately reproduce the dependence of the mode frequencies on magnetic field and reveal the spatial profiles of the excitations in the FL and RL. The FL and RL modes generate rectified voltage signals of opposite polarity, which makes this device a promising candidate for a tunable dual-frequency microwave signal detector. The simulations show that weak interlayer exchange coupling within the SAF enhances the mode amplitudes. We also calculate the response of the detector as a function of in-plane magnetic field bias and find that its sensitivity significantly increases with increasing field. We experimentally confirm this prediction via ST-FMR measurements as a function of in-plane magnetic field. Our results provide deeper understanding of quantized spin wave eigenmodes in nanoscale MTJs with perpendicular magnetic anisotropy and demonstrate the potential of these devices for frequency-selective dual-channel microwave signal detectors.

## I. INTRODUCTION

Further progress in spintronic technologies such as nonvolatile memories [1–3], magnetic sensors [4–7], microwave signal processing devices [8,9] and circuits for unconventional computing [10–12] relies on improvements of their core building blocks – nanoscale magnetic tunnel junctions (MTJs). In particular, MTJ-based spin-torque nano-oscillators (STNOs) and spin-torque diodes show characteristics (tunability, compactness, CMOS-compatibility, etc.) highly desirable for many applications [11,13] including microwave amplification [14] and detection [15,16], energy harvesting [17], and computing [18,19].

A powerful tool for characterization of nanoscale MTJs is analysis of the quantized spin wave eigenmodes in these structures. The main method of measuring the MTJ eigenmodes is spin-torque ferromagnetic resonance (ST-FMR) based on the spin-torque diode (STD) effect [11,20], which allows quantification of the eigenmode frequencies and damping. Such characterization is used for optimization of the material and device parameters for the targeted applications and evaluation of the device reliability via rapid screening of device-to-device variations [21,22]. Many applications utilizing MTJs, such as memories, require that one of the ferromagnetic layers within the MTJ is strongly pinned (reference layer - RL), and hence the excitation of a single main mode localized only in the MTJ free layer (FL) is expected. However, recent works have shown the excitation of non-uniform modes [23] or several linear [24–26] and non-linear [27–29] modes that have been attributed to both the FL and the RL of the MTJ, including devices having synthetic antiferromagnets (SAFs) as RLs [27,30–32].

Here, we unravel the dynamic properties of MTJs by combining experimental data from ST-FMR measurements with a full micromagnetic approach able to model the coupled dynamics of the FL and RL. We detect two modes characterized by opposite signs of the frequency dependence on the out-of-plane magnetic field for the antiparallel configuration of the MTJ. Our micromagnetic simulations reveal that one of the modes arises from the FL magnetization dynamics while the other mode belongs to the RL. The opposite polarity of the rectified voltage for the two modes as well as the mode tunability by a bias field make such MTJs promising candidates for a tunable dual-band microwave detector.

The paper is organized as follows. Sec. II presents the MTJ devices and ST-FMR measurements, Sec. III describes the experimental results, Sec IV presents the micromagnetic model, and Sec. V discusses the simulation results. Finally, Sec. VI summarizes the main conclusions of this study.

## II.  DEVICE DESCRIPTION AND MEASUREMENTS

MTJs were patterned into stadium-shaped nanopillars from magnetic multilayers of the following composition: bottom electric lead | SAF | tunnel barrier spacer | FL | top electric lead, where tunnel barrier spacer ≡ MgO (0.8 nm) , FL ≡ CoFeB (2.45 nm), SAF ≡ bottom SAF ferromagnet (1.67 nm, bSAF) | nonmagnetic metal spacer (0.41 nm) | top SAF ferromagnet (1.1 nm, tSAF) (see Fig. 1(a) for the sketch of the device). All ferromagnetic layers are designed to have perpendicular magnetic anisotropy (PMA) sufficient to ensure the equilibrium direction of magnetization along the layer normal [33]. The multilayers were deposited by magnetron sputtering. The data shown in this paper is from devices having the largest and smallest lateral dimensions of 65 and 30 nm in the film (x-y) plane, respectively (see Fig. 1 (a)). Fig. 1(b) shows an example of magnetoresistance scan as a function of the out-of-plane (z-axis) magnetic field. The RL of the device is set with the bSAF along the positive z-axis ($+\hat{z}$) and the tSAF along ($-\hat{z}$). The measured low ($R_p$) and high ($R_{ap}$) resistance states are $R_p = 7.5$ kΩ and $R_{ap} = 12$ kΩ, resulting in a tunnel magnetoresistance (TMR) ratio $TMR = (R_{ap} - R_p)/R_p$ of approximately 65%. The gradual decrease of the resistance in the high field regime ($H_{ext} > 300$ mT) of the AP state is due to the RL magnetization rotation. In addition, the small off-centering shift ($H_{shift} \sim 5$ mT) of the hysteresis loop is due to the stray magnetic field from the non-fully compensated SAF RL.

We use ST-FMR [34–36] to measure spin wave eigenmodes in the MTJ samples. Microwave current in the 2 – 26.5 GHz frequency range is applied directly to the sample through the microwave port of a bias tee, and a rectified voltage generated by the MTJ via the STD effect is measured at DC port of the bias tee. Resonances in the measured rectified signal correspond to the quantized spin wave eigenmodes of the MTJ.

We employ field-modulated ST-FMR detection [37]. In this variant of ST-FMR, the microwave power is fixed while the applied magnetic field is modulated, and a field-derivative of the rectified voltage is measured by a lock-in amplifier. The field-modulated signal shows improved signal-to-noise ratio allowing for more precise measurements of the frequency and linewidth of spin wave resonances [38].

The insets of Fig. 1(b) show two examples of ST-FMR data measured as a function of out-of-plane magnetic field $H_{ext} = \pm 100$ mT in the P and the AP states of the MTJ as indicated in the figure. We observe the ST-FMR signal amplitude for the AP state to be higher than that measured for the P state. In the rest of this paper, we only consider ST-FMR data measured in the AP state. We use vibrating sample magnetometry (VSM) to measure the saturation magnetizations of the FL, tSAF, and bSAF: $M_{S,FL} = 0.758$ MA/m, $M_{S,tSAF} = 0.950$ MA/m and $M_{S,bSAF} = 0.670$ MA/m.

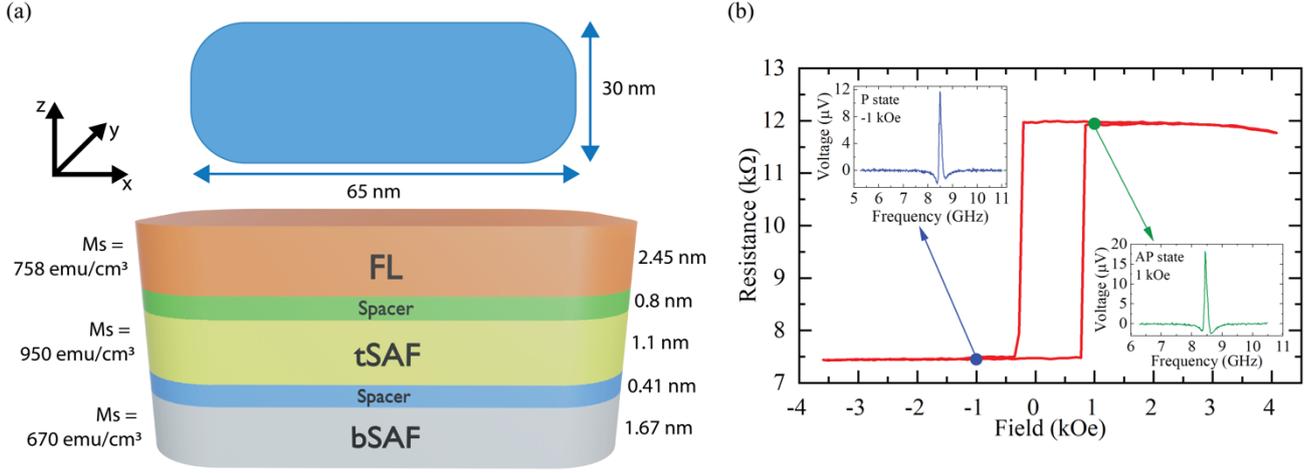

Fig. 1: **Measurement setup and MTJ characterization.** (a) Sketch of the studied stadium-shaped perpendicular MTJ. (b) MTJ resistance as a function of the out-of-plane magnetic field $H_{ext}$. Insets: ST-FMR response of the MTJ at the out-of-plane magnetic fields of -100 mT (blue) and +100 mT (green), corresponding to the parallel (P) and antiparallel (AP) configurations, respectively.

## III. EXPERIMENTAL RESULTS

A systematic characterization of spin wave eigenmodes in the AP state as a function of the out-of-plane field $H_{ext}$ is shown in Fig. 2, which displays ST-FMR spectra in the 7 – 26.5 GHz frequency range. Two modes are visible at large fields ($H_{ext} > 220$ mT). One of the modes (mode-1) is characterized by a blue frequency shift with $H_{ext}$, while the other mode (mode-2) exhibits a red shift. These results imply that mode-1 is related to the FL dynamics since we have considered the AP ground state (positive $H_{ext}$ is in the same direction as the FL magnetization), while mode-2 derives from the RL dynamics. Thus, in the following, we will refer to mode-1(2) as FL (RL) modes.

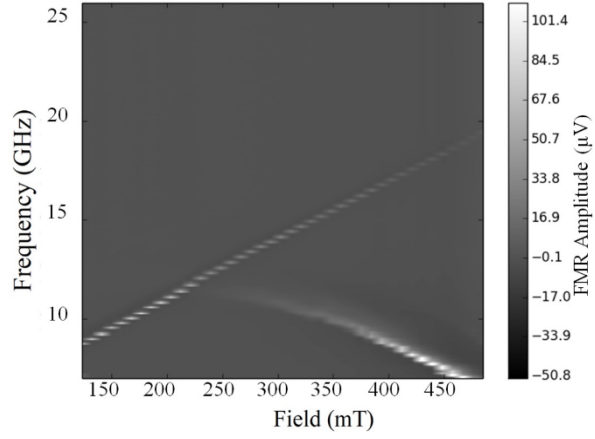

Fig. 2: Experimental results. Experimental ST-FMR spectra as a function of frequency and out-of-plane magnetic field.

The existence of two distinct modes can lead to several useful applications. First, the fact that the two modes are excited in two different magnetic elements of the MTJ enables separate characterization of the FL and RL layers and optimization of their parameters via a characterization/material optimization feedback loop. Second, the resonance frequencies of the two modes are separated by several GHz, allowing for realization of a dual-band spintronic diode, similarly to what already proposed in a previous work [29]. There two key differences between Ref. [29] and the present work. First, equilibrium magnetization direction of the FL and RL in Ref. [29] are in the sample plane, which limits the maximum frequency (S-band (2 – 4 GHz) and C-band (4 – 8 GHz) and X-band (8 – 12 GHz)) achievable by the detector. In contrast, Fig. 2 shows that the dual frequency detector based on a perpendicular MTJ can operate in the X-band (8 – 12 GHz) and $K_u$-band (12 – 18 GHz) used in satellite communications. Second, the two modes used in Ref. [29] are spatially non-uniform, which limits the maximum amplitude of the rectified signal generated by these modes. Our micromagnetic simulations described below reveal that the two excited modes in the perpendicular MTJ are spatially uniform allowing for efficient conversion of the mode dynamics into rectified voltage. We also show that application of an in-plane magnetic field to the perpendicular MTJ detector strongly increases the rectified signal without changing the spatial uniformity of the excited modes.

## IV. MICROMAGNETIC MODELING

To corroborate the interpretation of the experimental results and to predict new device functionalities, we perform micromagnetic simulations based on the numerical integration of the Landau-Lifshitz-Gilbert (LLG) equation by means of an in-house native GPU solver, PETASPIN [39,40]:

$$\frac{d\bm{m}}{d\tau} = -\frac{1}{1+\alpha_G^2}[(\bm{m}\times\bm{h}_{\text{eff}}) + \alpha_G \bm{m}\times(\bm{m}\times\bm{h}_{\text{eff}})] \quad (1)$$

where $\alpha_G$ is the Gilbert damping constant, $\bm{m} = \bm{M}/M_s$ is the normalized magnetization vector of the magnetic layer of saturation magnetization $M_s$, $\tau = \gamma_0 M_s t$ is the dimensionless time, $\gamma_0$ is electron gyromagnetic ratio and $\bm{h}_{\text{eff}} = \bm{h}_{\text{eff}}/(\mu_0 M_s)$ is the normalized effective field acting on the MTJ. $\bm{h}_{\text{eff}}$ includes the exchange ($\bm{h}_{\text{ex}}$), magnetostatic ($\bm{h}_{\text{dmag}}$), anisotropy ($\bm{h}_{\text{K}}$) and external ($\bm{h}_{\text{ext}}$) fields.

The MTJ is modeled as a 65 × 30 nm² stadium-shaped stack comprising three magnetic layers, FL, top SAF (tSAF), and bottom (bSAF), of thickness 2.45, 1.10 and 1.70 nm respectively, and two spacers of thickness 0.8 and 0.4 nm following sequence bottom bSAF/ spacer 1/ tSAF/ spacer 2/ FL. The SAF is characterized by the interlayer exchange coupling (IEC), and the interlayer effective field is $\bm{h}_{\text{IEC},i} = \frac{A_{\text{iex}}}{\mu_0 M_{s,i}^2 t_{\text{NM}}}\bm{m}_j$ [41], where $i,j$ are the indices of the tSAF and bSAF layers, $A_{\text{iex}}$ is the IEC constant, and $t_{\text{NM}}$ is the thickness of the spacer. IEC describes the Ruderman–Kittel–Kasuya–Yosida (RKKY) interaction between the local moments and the conduction electrons [42,43], which results in a coupling between the two ferromagnetic layers that oscillates between ferromagnetic and antiferromagnetic depending on the thickness of the spacer. For $t_{\text{NM}} = 0.4$ nm, as in our MTJ, the coupling is antiferromagnetic. The three magnetic layers are also coupled via the magnetostatic field. For the excitation of the magnetization dynamics in both the FL and RL, we rely upon the spin-transfer torque (STT) mechanism. This includes the effect of the back torque [44] on the tSAF magnetization due to the electrons scattered at the FL interface, induced by an AC current density $J_{ac} = J_{ac,0} \sin(2\pi f_{ac} t)$ of frequency $f_{ac}$ and amplitude $J_{ac,0}$. The STT term is added to Eq. (1) as:

$$\begin{cases} \boldsymbol{\tau}_{STT,FL} = \dfrac{g\mu_B}{eM_{s,FL}d_{FL}} J_{ac}[\boldsymbol{m}_{FL} \times (\boldsymbol{m}_{FL} \times \boldsymbol{m}_{tSAF}) - \alpha_G(\boldsymbol{m}_{FL} \times \boldsymbol{m}_{tSAF})] \\ \boldsymbol{\tau}_{STT,tSAF} = \dfrac{g\mu_B}{eM_{s,tSAF}d_{tSAF}}(-J_{ac})[\boldsymbol{m}_{tSAF} \times (\boldsymbol{m}_{tSAF} \times \boldsymbol{m}_{FL}) - \alpha_G(\boldsymbol{m}_{tSAF} \times \boldsymbol{m}_{FL})] \end{cases} \quad (2),$$

where $\boldsymbol{m}_{FL,tSAF}$ are the normalized magnetizations of FL and tSAF, $d_{FL,tSAF}$ are the thickness of FL and tSAF, $\mu_B$ is the Bohr magneton, $g$ is the Landè factor and $e$ the electron charge. $\boldsymbol{\tau}_{STT,FL}$ describes the STT acting on the FL, $\boldsymbol{\tau}_{STT,tSAF}$ represents the back torque exerted on the tSAF due to the reflected electrons at the FL interface. The opposite direction of those electrons is taken into account via the minus sign of $(-J_{ac})$ in $\boldsymbol{\tau}_{STT,tSAF}$. It follows that the two torques are characterized by a 180 degree phase difference.

In the simulations, we use the experimentally measured value of the FL magnetization ($M_s = 758$ kA/m) for all the magnetic layers. The FL and tSAF perpendicular anisotropy energy constants are extracted from the experimental frequency response shown in Fig. 2(a). In particular, by using Kittel's equation [20] for the two modes, we extract the resonance frequency $f_{Kit} = \dfrac{\gamma_0}{2\pi}\left(H_{ext} + H_{K,eff} + H_{shift}\right)$ by fitting the experimental frequency vs field curves. For the tSAF, we apply a linear approximation and verify numerically that we obtain frequencies similar to the experimental ones. Here $H_{K,eff}$ is the effective magnetic anisotropy field, from which we obtain $K_{U,FL} = 0.37$ MJ/m$^3$ and $K_{U,tSAF} = 0.6$ MJ/m$^3$ for the FL and tSAF, respectively. For the bSAF we set $K_{U,bSAF} = 1.0$ MJ/m$^3$, a value that makes the layer stable under large perpendicular fields.

The exchange stiffness ($A_{ex}$) and spin polarization ($\eta$) are taken from literature for similar systems: $A_{ex,FL} = A_{ex,tSAF} = A_{ex,bSAF} = 15$ pJ/m [45], $\eta = 0.66$ [45], while the Gilbert damping $\alpha_G = 0.006$ is extracted from experimental measurements of the FL mode linewidth as a function of resonance frequency [46], and we utilize this value for all the magnetic layers. To identify the IEC constant $A_{iex}$ for the tSAF | spacer | bSAF stack, we perform static simulations starting from an AP

state, i.e. bSAF and FL magnetizations in the positive z-direction ($+\hat{z}$), whereas tSAF in the negative z-direction ($-\hat{z}$). We systematically applied a positive out-of-plane external field, and observed that, with $A_{\text{iex}} = -0.1$ mJ/m$^2$, the tSAF magnetization remains stable up to 400 mT, in agreement with the experimental hysteresis loop of Fig. 1(b). Therefore, this value $A_{\text{iex}} = -0.1$ mJ/m$^2$ has been used, while qualitative similar results have been observed for simulations with larger $A_{\text{iex}}$. The device is discretized into cuboidal cells of dimensions 2.5 nm x 2.5nm x 0.4 nm, and all simulations are performed at zero temperature.

The rectified voltage $V_{\text{dc}}$, due to magnetization oscillations of both FL and tSAF, is computed as the time-averaged value of $V = J_{\text{ac},0} A \sin(\omega t) (R_p + \frac{1}{2}(R_{ap} - R_p)(1 - \boldsymbol{m}_{FL} \cdot \boldsymbol{m}_{tSAF}))$, where $J_{\text{ac},0} A \sin(\omega t)$ is the input current and $(R_p + \frac{1}{2}(R_{ap} - R_p)(1 - \boldsymbol{m}_{FL} \cdot \boldsymbol{m}_{tSAF}))$ is the oscillating magnetoresistance.

## V. RESULTS AND DISCUSSION
### A. Comparison of experimental data and micromagnetic calculations

To achieve a deeper understanding of the experimental data, we simulate the magnetization dynamics excited by a weak microwave current of magnitude $J_{\text{ac},0} = 0.5$ MA/cm$^2$ as a function of the out-of-plane field $\boldsymbol{H}_{\text{ext}} = H_{\text{ext}}\hat{z}$. In the simulations, the FL and bSAF are initialized with magnetization aligned along $+\hat{z}$, while the tSAF is in the antiparallel configuration ($-\hat{z}$). The system is first relaxed in the absence of the microwave current to an equilibrium condition. Subsequently, the microwave current is injected, and we compute the full frequency response as a function of the $\boldsymbol{H}_{\text{ext}}$, to compare it with the experimental data in Fig. 2. These simulations reveal excitation of two modes in the experimentally relevant frequency range. Fig. 3(a) shows the experimental (dashed lines extracted from the data in Fig. 2) and simulated (symbols) frequencies of the modes as a function of $H_{\text{ext}}$. We find a very good quantitative agreement between the experimental and simulated frequencies. Fig. 3(b) shows the simulated microwave detector response $V_{\text{dc}}(f)$ for $H_{\text{ext}} = 350$ mT that exhibits opposite polarity of $V_{\text{dc}}$ for the two modes, in agreement with the behavior observed in Ref. [29].

The micromagnetic simulations allow us to understand a number of features experimentally observed in this system. First, by performing a layer-by-layer analysis, we confirm that the mode-1 is generated in the FL, while mode-2 is excited in the RL. In particular, by means of this analysis we can identify this mode with the excitation of the tSAF, since this is more than two orders of magnitude larger than those observed in the bSAF. Second, we identify the origin of the frequency dependence of the modes: mode-1 (2) exhibits a blue (red) shift because the magnetization is oriented parallel (antiparallel) to the applied field, leading to a reduction (increase) of the precessional orbit. Hence, in the following mode-2 will be referred to as tSAF mode. Third, we confirm that the opposite polarity of $V_{dc}$ in Fig. 3(b) for the two modes arises from the 180 degree phase shift between the FL and tSAF magnetizations due to the back torque, as seen in Eq. 2.

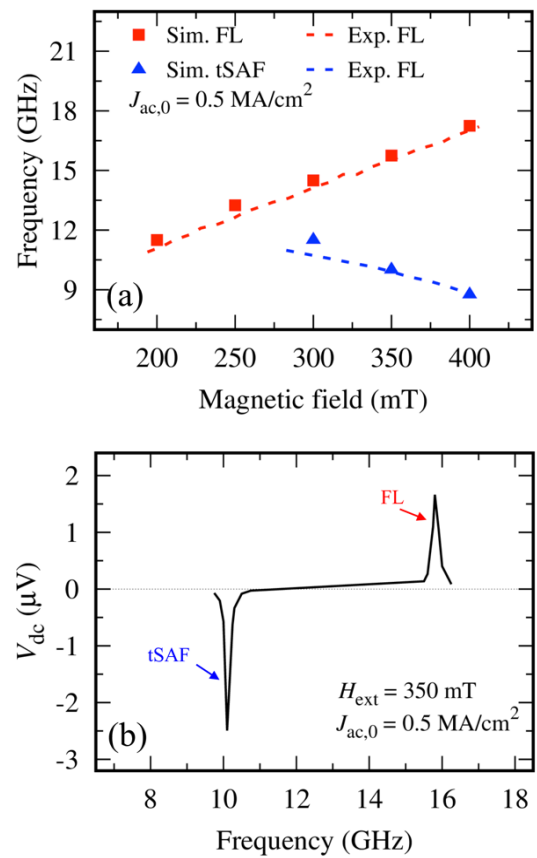

Fig. 3: Comparison between experiments and simulations. (a) ST-FMR resonance frequency vs field response comparing experiments (dashed lines) and simulations (red squares for the FL and blue

triangles for the RL). (b) Rectification curve showing $V_{dc}$ as a function of frequency for $H_{ext} = 350$ mT.

We have performed a systematic micromagnetic investigation of the microwave current density, Gilbert damping and interlayer exchange coupling strength on the resonant properties of the MTJ. The dependence on $J_{ac}$ exhibits the expected quadratic behavior and is discussed in more detail in Note S1 of Supplemental Material [47], while the linewidth of the $V_{dc}$ peaks at resonance increases linearly with $\alpha_G$, a result expected for FMR dynamics in these systems, see Note S2 of Supplemental Material [47]. In the following, we discuss in more detail the effect of interlayer exchange coupling.

### B. Effect of the interlayer exchange coupling

We vary $A_{iex}$ between $-0.1$ and $-1.5$ mJ/m$^2$ and study its impact on $V_{dc}(f)$. For this systematic analysis, we fix $J_{ac,0} = 2.5$ MA/cm$^2$. The main effect is the increase of tSAF resonance frequency with $|A_{iex}|$, while the FL resonance remains constant, as shown in Fig. 4(c). In addition, $V_{dc}$ amplitude at the resonance for the two modes increases for lower $|A_{iex}|$ (Fig. 4(d)). This is expected for the tSAF and leads to an enhancement of the STT acting on the FL, which, consequently, increases its precessional orbit and its frequency. This systematic study yields two main conclusions: (i) ST-FMR measurement is an easy and accessible tool to characterize the interlayer coupling of a SAF system, and (ii) for applications of this type of MTJ as a dual band detector, it is more desirable to use SAF with a smaller value of $|A_{iex}|$.

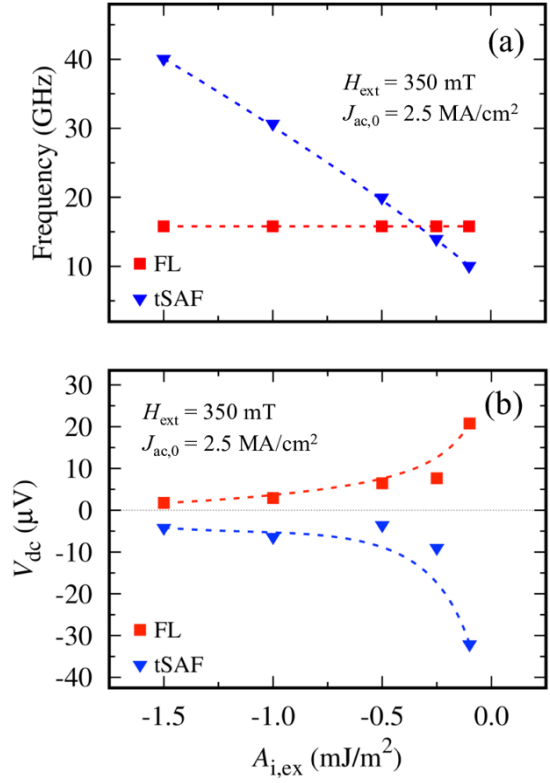

Fig. 4: Impact of interlayer exchange coupling for $H_{ext} = 350$ mT and $J_{ac,0} = 2.5$ MA/cm$^2$. (a) Resonance frequency dependence of the FL and tSAF modes on the interlayer exchange coupling ($A_{iex}$), and (b) dependence of $V_{dc}$ at resonance on $A_{iex}$. Lines are guides to the eye.

### C. In plane field ST-FMR response

So far, we have presented results for an out-of-plane $H_{ext}$. In this section, we study the MTJ frequency response as a function of an in-plane field applied in the x-direction (long axis of the stadium shape), and we fix $J_{ac,0} = 0.5$ MA/cm$^2$. Such an in-plane field introduces misalignment between directions of the FL and tSAF, which leads to a more efficient conversion of magnetic oscillations into resistance oscillations and thereby to a higher rectification efficiency.

Fig. 5(a) depicts the resonance frequency versus in-plane magnetic field. We identify three regions: region I where all magnetic layers (FL, tSAF, and bSAF) have an out-of-plane component of magnetization; region II where the FL is in-plane, while the tSAF and bSAF have an out-of-plane component of magnetization; region III where the FL and tSAF are in-plane, while the bSAF retains an out-of-plane component of magnetization. In region I, the FL starts to tilt towards the in-plane direction, hence the mode frequency decreases with the field, while the tSAF mode frequency is nearly constant (the field is smaller than the interlayer exchange field and does not affect the tSAF oscillation). In region II, the FL mode frequency follows the expected Kittel behavior for an in-plane ferromagnet under an in-plane field. In contrast, the tSAF magnetization continues to tilt towards the in-plane direction, thus the mode frequency decreases with the field. In region III, the FL keeps behaving as in region II, and the tSAF frequency is non-monotonic because the tSAF undergoes a transition from out-of-plane to in-plane [30,48]. In addition, a third mode (tSAF-2 mode) is excited for $H_{ext} \geq 800$ mT, and we identify it as a higher order tSAF layer mode.

Fig. 5(b) shows three rectification curves $V_{dc}(f)$ for $H_{ext} = 100, 500$ and 800 mT. $V_{dc}(f)$ has positive polarity in the FL mode and negative polarity for the tSAF-1 mode. In addition, for $H_{ext} = 800$ mT, $V_{dc}(f)$ exhibits three peaks, with the polarity of the tSAF-2 being negative. This suggests that the third mode is excited in the tSAF layer. Finally, Fig. 5(e) illustrates the spatial distribution of the tSAF-2 mode power in the FL and tSAF layer. The former is uniform, while the latter is non-uniform due to the combined effect of the non-uniform stray field and interlayer coupling with out-of-plane bSAF, as shown by the black curve in Fig. 5(b).

We wish to highlight that our MTJ with an in-plane field can be also used as a promising dual-band rf detector. Fig. 5 (c) shows the $V_{dc}$ amplitude at resonance as a function of the in-plane field. For fields in region I, the system exhibits a response spaced 15-20 GHz apart and the rectified signal has opposite polarities at the different peaks. In region II, $V_{dc}$ amplitude of tSAF exhibits a non linear behavior as a function of the field: $V_{dc}$ amplitude increases as the tSAF tilts towards the in-plane field, until the tilting angle exceeds 45 degree. Differently, the frequency of both modes decreases, as seen in Fig. 5 (a). Moreover, we wish to highlight that $V_{dc}$ amplitude increases of more than two order of magnitude with respect to the out-of-plane case. Therefore, our results suggest that the optimal working point of the detector is across region I and II.

Fig. 5 (d) shows our ST-FMR experimental data as a function of in-plane magnetic field for a 60 × 150 nm$^2$ stadium-shaped MTJ with the same magnetic multilayer structure as the device in Fig. 1 (a). The data reveals the lowest-frequency spin wave of the FL as well as several higher order FL modes in the frequency range used in this measurement. As predicted by our micromagnetic simulations, the amplitude of the rectified signal strongly increases with increasing magnitude of the in-plane magnetic field in the relevant region I.

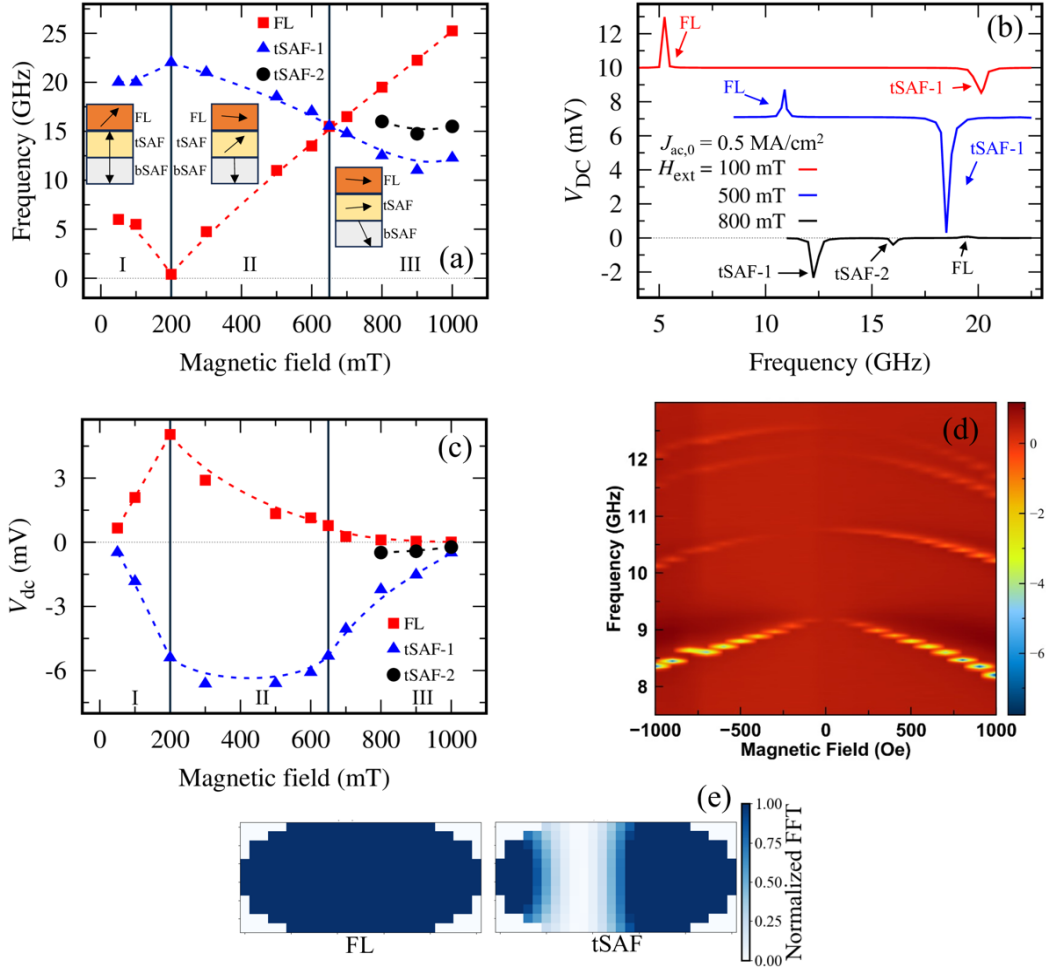

Fig. 5: In-plane field ST-FMR response for $J_{ac,0} = 0.5$ MA/cm$^2$. (a) ST-FMR resonance frequency vs field response for in-plane external field obtained by micromagnetic simulations. Lines are guide to the eye; insets represent sketch of the magnetic configuration of the magnetic layers. (b) Rectification curves for $H_{ext} = 100, 500, 800$ mT, where curves are shifted along the y-axis for better readability. (c) $V_{dc}$ amplitude at resonance as a function of in-plane field; lines are a guide to the eye.

(d) Experimental ST-FMR spectra as a function of frequency and in-plane magnetic field for a 60 × 150 nm² stadium-shaped MTJ. (e) Spatial distribution of the tSAF-2 mode in the FL and tSAF layer for $H_{\text{ext}} = 800$ mT; each plot is normalized with respect to its maximum FFT amplitude (color palette: 0 – white, 1 – dark blue).

The intersection of the FL and tSAF-1 modes at 650 mT in Fig. 5(a) creates potential for collective mode dynamics and mode hybridization with an enhancement of the resulting $V_{\text{dc}}$. However, below we show that this does not occur in our system. Fig. 6 (a) depicts the time traces of $m_{\text{x,FL}}$ and $m_{\text{x,tSAF}}$ at 15.5 GHz under $H_{\text{ext}} = 650$ mT (note that we have multiplied $m_{\text{x,FL}}$ by a factor 100 for readability, which also demonstrates the larger amplitudes of the oscillations in the tSAF than in the FL). Even though the two modes are excited at the same frequency, they interact destructively because the magnetizations are nearly out-of-phase, as shown in Fig. 6(b). This phase difference is due to the opposite sign of the torque acting on the magnetizations of the two layers. Therefore, the modes cannot hybridize. Fig. 6(b) shows the $V_{\text{dc}}$ amplitude at the tSAF-1 mode for $H_{\text{ext}} = 600, 650$ and 700 mT. $V_{\text{dc}}$ decreases with field with no enhancement around 650 mT, confirming that there is no collective dynamics. Therefore, our results demonstrate that if we wish to enhance sensitivity of the detector due to mode hybridization, an alternative MTJ configuration maximizing the mode coupling is needed [49].

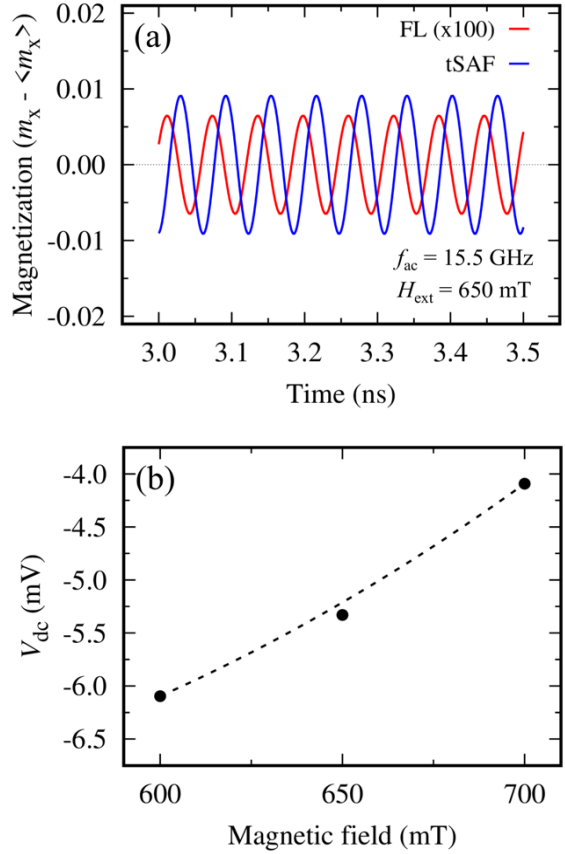

Fig. 6: Study of collective dynamics. (a) Time traces of the x-component of the magnetization of FL and tSAF at 15.5 GHz and $H_{ext} = 650$ mT, where the FL magnetization has been multiplied by a factor 100 for readability. (b) Plot of $V_{dc}$ amplitude for $H_{ext} = 600, 650, 700$ mT at the respective tSAF-1 frequency. All data is obtained with $J_{ac,0} = 0.5$ MA/cm$^2$.

### D. A comparison with previous works

The existence of two magnetic modes has been analyzed in extended SAF multilayers [27,30–32], and was explained in terms of acoustic and optical excitation modes as in antiferromagnets [50,51]. However, less attention has been paid to the spin wave mode structure in nanoscale MTJs. Specifically, in perpendicular MTJs, these modes have been primarily studied in the regime of out-out-plane fields where the frequencies of both modes vary linearly with the applied field [25]. In in-plane MTJs, which have been more extensively investigated in literature, mode-2 exhibits different qualitative behaviors as a function of field [24,27–29]. In details, Chao *et al* [24] find two modes, one characterized by a monotonic increase of the frequency with the field, the other by a frequency reduction at low magnetic fields followed by a frequency increase at high magnetic fields. Helmer *et al* [27] find several modes exhibiting a similar frequency trend for fields along the easy-axis, increasing for small fields and decreasing for larger fields. Zhang *et al* [28] study only small fields and find two modes, one with an increasing frequency and the other decreasing with field. Finally, Gupta *et al* [29] measure two modes, one with an increasing frequency with field and the other with a nearly flat dispersion.

Regarding perpendicular MTJs, Lyu *et al* [25] study an MTJ with a SAF as FL, with large in-plane dimensions and focus on the mode structure in the parallel configuration of the MTJ, where magnetizations of all the layers within the device are aligned parallel to the applied field. Differently, here (*i*) we investigate MTJs with sub-100 nm in-plane size and (*ii*) the SAF structure makes the RL. As RL, the SAF is characterized by a larger anisotropy and, thus, the SAF is able to retain an antiparallel alignment under an external field, which results in the dual mode response described above. We also note that a similar dual mode response, to that described in our work, was observed in Ref. [22].

## VI. SUMMARY

In conclusion, our combined experimental and theoretical investigation of quantized spin wave eigenmodes in perpendicular magnetic tunnel junctions (MTJs) provides new insights into the coupled magnetization dynamics of the free layer (FL) and synthetic antiferromagnet (SAF) reference layer (RL). Spin-torque ferromagnetic resonance (ST-FMR) measurements reveal the excitation of two distinct spin wave modes, with opposite frequency shifts in response to an applied out-of-plane magnetic field. Our micromagnetic simulations successfully replicate the observed mode properties and reveal opposite rectified signal polarities for the two modes. This characteristic behavior positions the device as a promising candidate for tunable dual-frequency microwave signal detection. The simulations also illustrate the spatial profiles of excitations in both layers and show that weak interlayer exchange coupling in the SAF enhances the mode amplitudes. Additionally, we find that the detector's sensitivity increases with an in-plane magnetic field bias. We also show that when magnetic field bias tunes the FL and RL mode frequencies to the same value, these modes do not hybridize due to their weak coupling in the perpendicular MTJ geometry. Our findings enhance the understanding of spin wave dynamics in nanoscale MTJs with perpendicular magnetic anisotropy and demonstrate their potential for application in frequency-selective dual-channel microwave detectors.

## ACKNOWLEDGEMENTS

AM, RT, MC, and GF thank the projects PRIN 2020LWPKH7 "The Italian factory of micromagnetic modelling and spintronics", PRIN20222N9A73 "SKYrmion-based magnetic tunnel junction to design a temperature SENSor—SkySens", funded by the Italian Ministry of Research, and the project number 101070287—SWAN-on-chip—HORIZON-CL4- 2021-DIGITAL EMERGING-01. AM, RT, MC, and GF are with the Petaspin TEAM and thank the support of the PETASPIN association (www.petaspin.com). AM, RT and MC acknowledge support from the Project PE0000021, "Network 4 Energy Sustainable Transition – NEST", funded by the European Union – NextGenerationEU, under the National Recovery and Resilience Plan (NRRP), Mission 4 Component 2 Investment 1.3 - Call for tender No. 1561 of 11.10.2022 of Ministero dell'Università e della Ricerca (MUR) (CUP C93C22005230007). CS and IK acknowledge funding and technical support by Samsung GRO Program. Funding from the National Science Foundation via awards ECCS-221369 and DMREF-232420 is also acknowledged.

## REFERENCES

[1] S. Bhatti, R. Sbiaa, A. Hirohata, H. Ohno, S. Fukami, and S. N. Piramanayagam, Spintronics based random access memory: a review, Materials Today **20**, 530 (2017).

[2] G. Siracusano, R. Tomasello, M. D'Aquino, V. Puliafito, A. Giordano, B. Azzerboni, P. Braganca, G. Finocchio, and M. Carpentieri, Description of Statistical Switching in Perpendicular STT-MRAM Within an Analytical and Numerical Micromagnetic Framework, IEEE Trans Magn **54**, 1400210 (2018).

[3] B. Jinnai, K. Watanabe, S. Fukami, and H. Ohno, Scaling magnetic tunnel junction down to single-digit nanometers—Challenges and prospects, Appl Phys Lett **116**, 160501 (2020).

[4] M. A. Khan, J. Sun, B. Li, A. Przybysz, and J. Kosel, Magnetic sensors-A review and recent technologies, Engineering Research Express **3**, 022005 (2021).

[5] B. Fang et al., Experimental Demonstration of Spintronic Broadband Microwave Detectors and Their Capability for Powering Nanodevices, Phys Rev Appl **11**, 014022 (2019).

[6] N. Maciel, E. Marques, L. Naviner, Y. Zhou, and H. Cai, Magnetic Tunnel Junction Applications, Sensors **20**, 121 (2019).

[7] Z. Q. Lei, G. J. Li, W. F. Egelhoff, P. T. Lai, and P. W. T. Pong, Review of Noise Sources in Magnetic Tunnel Junction Sensors, IEEE Trans Magn **47**, 602 (2011).

[8] S. Louis, O. Sulymenko, V. Tiberkevich, J. Li, D. Aloi, O. Prokopenko, I. Krivorotov, E. Bankowski, T. Meitzler, and A. Slavin, Ultra-fast wide band spectrum analyzer based on a rapidly tuned spin-torque nano-oscillator, Appl Phys Lett **113**, 112401 (2018).

[9] P. Y. Artemchuk, J. Zhang, O. V. Prokopenko, E. N. Bankowski, T. J. Meitzler, I. N. Krivorotov, J. A. Katine, V. S. Tyberkevych, and A. N. Slavin, Measurement of Microwave Signal Frequency by a Pair of Spin-Torque Microwave Diodes, IEEE Magn Lett **12**, 4502205 (2021).

[10] Z. Zeng, G. Finocchio, and H. Jiang, Spin transfer nano-oscillators, Nanoscale **5**, 2219 (2013).

[11] G. Finocchio, R. Tomasello, B. Fang, A. Giordano, V. Puliafito, M. Carpentieri, and Z. Zeng, Perspectives on spintronic diodes, Appl Phys Lett **118**, 160502 (2021).

[12] N. A. Aadit, A. Grimaldi, M. Carpentieri, L. Theogarajan, J. M. Martinis, G. Finocchio, and K. Y. Camsari, Massively parallel probabilistic computing with sparse Ising machines, Nat Electron **5**, 460 (2022).

[13] T. Chen, R. K. Dumas, A. Eklund, P. K. Muduli, A. Houshang, A. A. Awad, P. Durrenfeld, B. G. Malm, A. Rusu, and J. Akerman, Spin-Torque and Spin-Hall Nano-Oscillators, Proceedings of the IEEE **104**, 1919 (2016).

[14] K. Zhu et al., Nonlinear amplification of microwave signals in spin-torque oscillators, Nat Commun **14**, 2183 (2023).

[15] D. Marković, N. Leroux, A. Mizrahi, J. Trastoy, V. Cros, P. Bortolotti, L. Martins, A. Jenkins, R. Ferreira, and J. Grollier, Detection of the Microwave Emission from a Spin-Torque Oscillator by a Spin Diode, Phys Rev Appl **13**, 044050 (2020).

[16] B. Fang et al., Experimental Demonstration of Spintronic Broadband Microwave Detectors and Their Capability for Powering Nanodevices, Phys Rev Appl **11**, 014022 (2019).

[17] S. Hemour and K. Wu, Radio-frequency rectifier for electromagnetic energy harvesting: Development path and future outlook, Proceedings of the IEEE **102**, 1667 (2014).

[18] J. Cai, L. Zhang, B. Fang, W. Lv, B. Zhang, G. Finocchio, R. Xiong, S. Liang, and Z. Zeng, Sparse neuromorphic computing based on spin-torque diodes, Appl Phys Lett **114**, 192402 (2019).

[19] L. Mazza, V. Puliafito, E. Raimondo, A. Giordano, Z. Zeng, M. Carpentieri, and G. Finocchio, Computing with Injection-Locked Spintronic Diodes, Phys Rev Appl **17**, 014045 (2022).

[20] A. A. Tulapurkar, Y. Suzuki, A. Fukushima, H. Kubota, H. Maehara, K. Tsunekawa, D. D. Djayaprawira, N. Watanabe, and S. Yuasa, Spin-torque diode effect in magnetic tunnel junctions, Nature **438**, 339 (2005).

[21] B. Dieny et al., Opportunities and challenges for spintronics in the microelectronics industry, Nat Electron **3**, 446 (2020).

[22] C. J. Safranski, Y.-J. Chen, I. N. Krivorotov, and J. Z. Sun, Material parameters of perpendicularly magnetized tunnel junctions from spin torque ferromagnetic resonance techniques, Appl Phys Lett **109**, 132408 (2016).

[23] L. Zhang, J. Cai, B. Fang, B. Zhang, L. Bian, M. Carpentieri, G. Finocchio, and Z. Zeng, Dual-band microwave detector based on magnetic tunnel junctions, Appl Phys Lett **117**, 072409 (2020).

[24] X. Chao, Y. Zhang, and J.-P. Wang, *Spin-Torque Oscillation Modes of a Composite Synthetic Antiferromagnetic Free Layer in Dual Magnetic Tunnel Junctions*, in *2021 IEEE International Magnetic Conference (INTERMAG)*, Vols. 2021-April (IEEE, 2021), pp. 1–5.

[25] D. Lyu, D. Zhang, D. B. Gopman, Y. Lv, O. J. Benally, and J.-P. Wang, Ferromagnetic resonance and magnetization switching characteristics of perpendicular magnetic tunnel junctions with synthetic antiferromagnetic free layers, Appl Phys Lett **120**, 012404 (2022).

[26] A. Sidi El Valli et al., Size-dependent enhancement of passive microwave rectification in magnetic tunnel junctions with perpendicular magnetic anisotropy, Appl Phys Lett **120**, 012406 (2022).

[27] A. Helmer, S. Cornelissen, T. Devolder, J.-V. Kim, W. van Roy, L. Lagae, and C. Chappert, Quantized spin-wave modes in magnetic tunnel junction nanopillars, Phys Rev B **81**, 094416 (2010).

[28] Y. Zhang, H. Zhao, A. Lyle, P. A. Crowell, and J.-P. Wang, Spin torque oscillation modes of a dual magnetic tunneling junction, J Appl Phys **109**, 07D307 (2011).

[29] P. Gupta, N. Sisodia, T. Bohnert, J. D. Costa, R. Ferreira, and P. K. Muduli, Dual Band Radio Frequency Detector Based on the Simultaneous Excitation of Free and Reference Layer in a Magnetic Tunnel Junction, IEEE Electron Device Letters **44**, 172 (2023).

[30] Z. Zhang, L. Zhou, P. E. Wigen, and K. Ounadjela, Using Ferromagnetic Resonance as a Sensitive Method to Study Temperature Dependence of Interlayer Exchange Coupling, Phys Rev Lett **73**, 336 (1994).

[31] A. Layadi, Study of the resonance modes of a magnetic tunnel junction-like system, Phys Rev B **72**, 024444 (2005).

[32] A. Layadi, Resonant and Switching Fields for a Weakly Coupled Magnetic Tunnel Junction System, SPIN **06**, 1640011 (2016).

[33] I. Barsukov, Y. Fu, A. M. Gonçalves, M. Spasova, M. Farle, L. C. Sampaio, R. E. Arias, and I. N. Krivorotov, Field-dependent perpendicular magnetic anisotropy in CoFeB thin films, Appl Phys Lett **105**, 152403 (2014).

[34] J. C. Sankey, P. M. Braganca, A. G. F. Garcia, I. N. Krivorotov, R. A. Buhrman, and D. C. Ralph, Spin-Transfer-Driven Ferromagnetic Resonance of Individual Nanomagnets, Phys Rev Lett **96**, 227601 (2006).

[35] J. N. Kupferschmidt, S. Adam, and P. W. Brouwer, Theory of the spin-torque-driven ferromagnetic resonance in a ferromagnet/normal-metal/ferromagnet structure, Phys Rev B **74**, 134416 (2006).

[36] L. Torres, G. Finocchio, L. Lopez-Diaz, E. Martinez, M. Carpentieri, G. Consolo, and B. Azzerboni, Micromagnetic modal analysis of spin-transfer-driven ferromagnetic resonance of individual nanomagnets, J Appl Phys **101**, 09A502 (2007).

[37] A. M. Gonçalves, I. Barsukov, Y.-J. Chen, L. Yang, J. A. Katine, and I. N. Krivorotov, Spin torque ferromagnetic resonance with magnetic field modulation, Appl Phys Lett **103**, 172406 (2013).

[38] I. Barsukov, H. K. Lee, A. A. Jara, Y.-J. Chen, A. M. Gonçalves, C. Sha, J. A. Katine, R. E. Arias, B. A. Ivanov, and I. N. Krivorotov, Giant nonlinear damping in nanoscale ferromagnets, Sci Adv **5**, eaav6943 (2019).

[39] A. Giordano, G. Finocchio, L. Torres, M. Carpentieri, and B. Azzerboni, Semi-implicit integration scheme for Landau–Lifshitz–Gilbert-Slonczewski equation, J Appl Phys **111**, 07D112 (2012).

[40] L. Lopez-Diaz, D. Aurelio, L. Torres, E. Martinez, M. A. Hernandez-Lopez, J. Gomez, O. Alejos, M. Carpentieri, G. Finocchio, and G. Consolo, Micromagnetic simulations using Graphics Processing Units, J Phys D Appl Phys **45**, 323001 (2012).

[41] E. Darwin, R. Tomasello, P. M. Shepley, N. Satchell, M. Carpentieri, G. Finocchio, and B. J. Hickey, Antiferromagnetic interlayer exchange coupled Co68B32/Ir/Pt multilayers, Sci Rep **14**, 95 (2024).

[42] M. A. Ruderman and C. Kittel, Indirect exchange coupling of nuclear magnetic moments by conduction electrons, Physical Review **96**, 99 (1954).

[43] Y. Yafet, Ruderman-Kittel-Kasuya-Yosida range function of a one-dimensional free-electron gas, Phys Rev B **36**, 3948 (1987).

[44] Z. Zeng, G. Finocchio, and H. Jiang, Spin transfer nano-oscillators, Nanoscale **5**, 2219 (2013).

[45] V. Krizakova, E. Grimaldi, K. Garello, G. Sala, S. Couet, G. S. Kar, and P. Gambardella, Interplay of Voltage Control of Magnetic Anisotropy, Spin-Transfer Torque, and Heat in the Spin-Orbit-Torque Switching of Three-Terminal Magnetic Tunnel Junctions, Phys Rev Appl **15**, 054055 (2021).

[46] G. D. Fuchs et al., Spin-torque ferromagnetic resonance measurements of damping in nanomagnets, Appl Phys Lett **91**, 062507 (2007).

[47] Supplemental Material, See Supplemental Material at [Link..] for a Systematic Study of the Effect of the Input Mircowave Current Density and Gilbert Damping on the Resonant Properties of the MTJ (n.d.).

[48]　T. Devolder, E. Liu, J. Swerts, S. Couet, T. Lin, S. Mertens, A. Furnemont, G. Kar, and J. De Boeck, Ferromagnetic resonance study of composite Co/Ni - FeCoB free layers with perpendicular anisotropy, Appl Phys Lett **109**, 142408 (2016).

[49]　A. Etesamirad, R. Rodriguez, J. Bocanegra, R. Verba, J. Katine, I. N. Krivorotov, V. Tyberkevych, B. Ivanov, and I. Barsukov, Controlling Magnon Interaction by a Nanoscale Switch, ACS Appl Mater Interfaces **13**, 20288 (2021).

[50]　M. Hagiwara, K. Katsumata, H. Yamaguchi, M. Tokunaga, I. Yamada, M. Gross, and P. Goy, A complete frequency-field chart for the antiferromagnetic resonance in MnF2, Int J Infrared Millimeter Waves **20**, 617 (1999).

[51]　V. Baltz, A. Manchon, M. Tsoi, T. Moriyama, T. Ono, and Y. Tserkovnyak, Antiferromagnetic spintronics, Rev Mod Phys **90**, 15005 (2018).

# SUPPLEMENTARY INFORMATION

# Spin wave eigenmodes in nanoscale magnetic tunnel junctions with perpendicular magnetic anisotropy – Supplemental material

Andrea Meo[1], Chengcen Sha[2], Emily Darwin[3], Riccardo Tomasello[1], Mario Carpentieri[1], Ilya N. Krivorotov[2], Giovanni Finocchio[4]

[1] *Department of Electrical and Information Engineering, Politecnico di Bari, I-70125 Bari, Italy*

[2] *Department of Physics and Astronomy, University of California, Irvine, California 92697, USA*

[3] *Empa, Swiss Federal Laboratories for Materials Science and Technology, CH-8600 Dübendorf, Switzerland*

[4] *Department of Mathematical and Computer Sciences, Physical Sciences and Earth Sciences, University of Messina, I-98166 Messina, Italy*

[*]Corresponding author: giovanni.finocchio@unime.it

## Supplementary Note S1

Fig. S1 shows the $V_{dc}$ amplitude at the resonance frequencies for $J_{ac,0}$ in the range 0.5 – 2.5 MA/cm$^2$. We fix $H_{ext} = 350$ mT, since the two modes are well separated in frequency at this field. The resonance frequencies of 15.8 GHz for the FL and 10.1 GHz for the tSAF do not change with $J_{ac}$ and $V_{dc}$ amplitude of the two modes increase quadratically with $J_{ac}$. In fact, the magnetoresistance of the MTJ depends on the relative angle between two magnetic layers, i.e. $\boldsymbol{m}_{FL} \cdot \boldsymbol{m}_{tSAF}$, that is linear in $J_{ac}$. Since $V_{dc}$ is proportional to the product of the latter and $J_{ac}$, it follows the quadratic behavior [1].

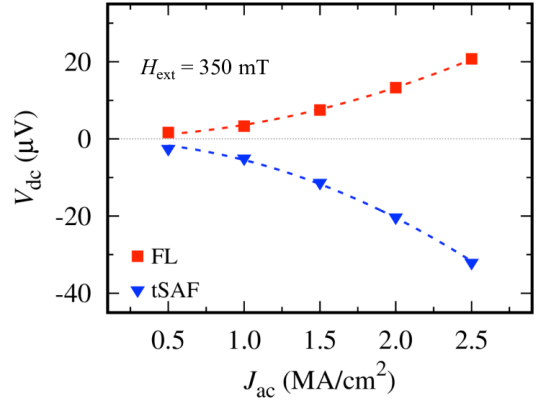

Fig. S1: Impact of microwave current density. Dependence of $V_{dc}$ at resonance on the injected microwave current density ($J_{ac}$) for $H_{ext} = 350$ mT. Lines represent a quadratic fit to the simulation data.

## Supplementary Note S2

To analyze the effect of Gilbert damping ($\alpha_G$), we consider both the $V_{dc}$ amplitude at the resonance frequency and the linewidth of the $V_{dc}$ peaks at resonance at fixed $J_{ac,0} = 0.5$ MA/cm$^2$. We fix $H_{ext} = 350$ mT, since the two modes are well separated in frequency at this field. Fig. S2(a) shows the dependence of the $V_{dc}$ amplitude at the resonance frequency for the FL and tSAF modes, from which we can see that as the damping increases both peaks amplitude decrease, as expected for a stiffer system. It is worth observing that for applications such as the dual band detector, where it is desirable to obtain a large output signal from both modes, layers characterized by a small damping should be chosen. Fig. S2(b) shows the resonance linewidth, extracted by fitting the simulated $V_{dc}(f)$ to a Lorentzian function [S1,S2], as a function of damping from $\alpha_G = 0.006$ (experimental value) to $\alpha_G = 0.05$. The linewidth is linear and increases with the damping, resulting in a broader rectification curve. This trend is expected for standard FMR in these systems [S2,S3] and is a further confirmation of the model.

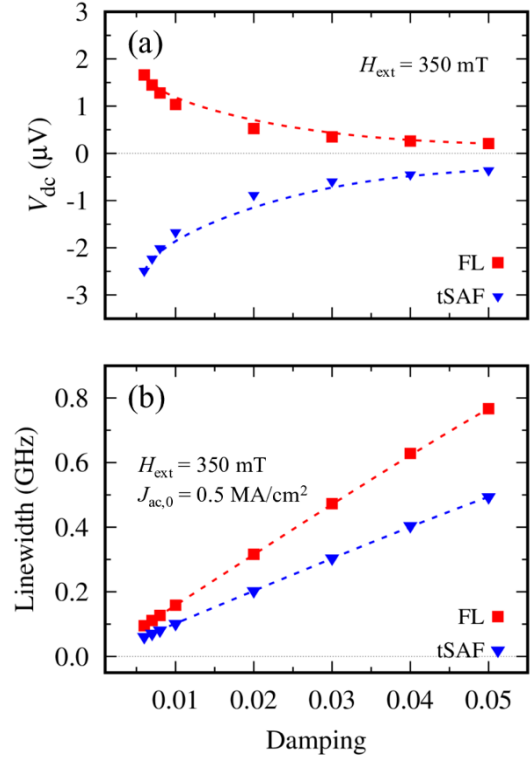

Fig. S2: Impact of Gilbert damping. (a) Dependence of the $V_{dc}$ amplitude at the resonance frequency for the FL (red) and tSAF (blue) modes; lines are guides to the eye. and (b) linewidth dependence of $V_{dc}$ at resonance on the Gilbert damping ($\alpha_G$). Simulations are performed with an input microwave current $J_{ac,0} = 0.5$ MA/cm² and under an eternal field $H_{ext} = 350$ mT. Lines represent the fit to a line.

# References

[S1] A. A. Tulapurkar, Y. Suzuki, A. Fukushima, H. Kubota, H. Maehara, K. Tsunekawa, D. D. Djayaprawira, N. Watanabe, and S. Yuasa, Spin-torque diode effect in magnetic tunnel junctions, Nature 438, 339 (2005).

[S2] J. C. Sankey, P. M. Braganca, A. G. F. Garcia, I. N. Krivorotov, R. A. Buhrman, and D. C. Ralph, Spin-Transfer-Driven Ferromagnetic Resonance of Individual Nanomagnets, Phys Rev Lett 96, 227601 (2006).

[S3] B. Heinrich, Y. Tserkovnyak, G. Woltersdorf, A. Brataas, R. Urban, and G. E. W. Bauer, Dynamic Exchange Coupling in Magnetic Bilayers, Phys Rev Lett 90, 187601 (2003).